\documentclass[aps,preprint,superscriptaddress]{revtex4}
\usepackage{graphicx}
\usepackage{amsmath}        
\usepackage{mathtools}        
\usepackage{setspace}
\usepackage[colorlinks]{hyperref}
\usepackage{color}
\definecolor{darkred}{rgb}{0.5,0,0}
\definecolor{darkgreen}{rgb}{0,0.5,0}
\definecolor{darkblue}{rgb}{0,0,0.5}
\hypersetup{ colorlinks,
linkcolor=darkblue,
filecolor=darkblue,
urlcolor=darkblue,
citecolor=darkblue }
\begin{document}

\bibliographystyle{apsrev}

\newcommand{\comment}[1]{}


\newcommand{\G}{{\stackrel{\leftrightarrow}{\bf G}}}
\newcommand{\SG}{{\stackrel{\leftrightarrow}{\bf S}}}
\newcommand\localizedplasmon{LSSP}
\newcommand\atomgeneral{\mathbf{X}}
\newcommand\atoma{\mathbf{x}}
\newcommand\atomb{\mathbf{y}}
\newcommand\bothatoms{\mathbf{eq}}
\newcommand\levela{\mathbf{0}}
\newcommand\levelb{\mathbf{1}}
\newcommand\levelc{\mathbf{2}}
\newcommand\levelintermediate{\mathbf{i}}
\newcommand\levelintermediateb{\mathbf{1i}}
\newcommand\levelintermediatec{\mathbf{2i}}
\newcommand\levelfinal{\mathbf{f}}
\newcommand\levelgenerala{\mathbf{m}}
\newcommand\levelgeneralb{\mathbf{l}}
\newcommand\levelgeneralc{\mathbf{n}}
\newcommand\otherlevela{\mathbf{x}}
\newcommand\otherlevelb{\mathbf{y}}
\newcommand\population[2]{\prescript{{#1}}{}{N^{#2}}}
\newcommand\partition[2]{\prescript{{#1}}{}{Z^{#2}}}
\newcommand\upconversion[3]{\prescript{{#1}}{}{\phi^{#2}_{#3}}}
\newcommand\planck{\mathbf{h}}
\newcommand\einsteinB[2]{\prescript{{#1}}{}{B^{#2}}}
\newcommand\energydensity[2]{\prescript{{#1}}{}{W^{#2}}}
\newcommand\cspeed{\mathbf{c}}
\newcommand\einsteinA[2]{\prescript{{#1}}{}{A^{#2}}}
\newcommand\absolute[1]{{|{#1}|}}
\renewcommand\time{\mathbf{t}}
\newcommand\radiative{\mathbf{R}}
\newcommand\nonradiative{\mathbf{NR}}
\newcommand\isolated{0}
\newcommand\epsilonglass{\epsilon_{glass}}
\newcommand\epsilonsilver{\epsilon_{Ag}}
\newcommand\total{T}
\newcommand\radiativeisolated{\radiative\isolated}
\newcommand\nonradiativeisolated{\nonradiative\isolated}
\newcommand\totalisolated{\total\isolated}

\newcommand\alllevels{All}
\newcommand\absorption{abs}
\newcommand\excitedstateabsorption{ESA}
\newcommand\nonradiativeenergytransfer{NRET}

\newcommand\local{local}
\newcommand\equilibrium{eq}
\newcommand\intensitydensity[1]{I_{#1}}
\newcommand\genericanglea{\theta_1}
\newcommand\genericangleb{\theta_2}
\newcommand\intensity{\mathbf{I}}
\newcommand\unitvector[1]{\mathbf{u_{#1}}}
\newcommand\intensitylocal[2]{^{#1}\intensity^{\local}_{#2}}

\newcommand\intensityout{P_{out}}
\newcommand\intensityoutisolated{P^{\isolated}}
\newcommand\enhancement[2]{{^{#1}K_{#2}}}
\newcommand\frequency[2]{{^{#1}\omega_{#2}}}
\newcommand\frequencyin{\frequency{}{in}}
\newcommand\frequencyout{\frequency{}{out}}

\newcommand\lifetime[3]{\prescript{{#1}}{}{\tau^{#3}_{#2}}}
\newcommand\quantumyield[3]{\prescript{{#1}}{}{\eta^{#3}_{#2}}}
\newcommand\constant{C_{te}}
\newcommand\decay[3]{\prescript{{#1}}{}{\Gamma^{#3}_{#2}}}
\newcommand\efield[3]{\prescript{{#1}}{}{E^{#3}_{#2}}}
\newcommand\decayheat[2]{\prescript{{#1}}{}{k^{NR}_{#2}}}

\newcommand\crosssection[3]{\prescript{{#1}}{}{\sigma^{#3}_{#2}}}

\newcommand{\Eq}[1]{Eq.~#1}
\newcommand{\Fig}[1]{Fig.~#1}
\newcommand{\Eqs}[1]{Eqs.~#1}
\newcommand{\Figs}[1]{Figs.~#1}
\newcommand{\Section}[1]{Sec.~#1}

\newcommand\exponent[1]{\mathbf{e^{#1}}}
\newcommand\bra{\mathbf{|}}
\newcommand\ket{\mathbf{>}}
\newcommand\evector{\vec{E}}
\newcommand\kvector{\vec{k}}

\newcommand\positionvector{\mathbf{\vec{1_d}}}
\newcommand\positionmod{\mathbf{{d}}}
\newcommand\nanometer{nm}
\newcommand\wavelength{\lambda}


\setstretch{1.5}

\title{Influence of metallic nanoparticles on upconversion processes}

\author{R. Esteban}
\affiliation{Laboratoire d'\'Energ\'etique Mol\'eculaire,
Macroscopique et Combustion,\\ \'Ecole Centrale Paris,
Centre National de la Recherche Scientifique, \\
Grande Voie des Vignes, 92295 Ch\^atenay-Malabry Cedex, France}
\author{M. Laroche}
\affiliation{Laboratoire d'\'Energ\'etique Mol\'eculaire,
Macroscopique et Combustion,\\ \'Ecole Centrale Paris,
Centre National de la Recherche Scientifique, \\
Grande Voie des Vignes, 92295 Ch\^atenay-Malabry Cedex, France}
\author{J.-J. Greffet}

\affiliation{Laboratoire d'\'Energ\'etique Mol\'eculaire,
Macroscopique et Combustion,\\ \'Ecole Centrale Paris,
Centre National de la Recherche Scientifique, \\
Grande Voie des Vignes, 92295 Ch\^atenay-Malabry Cedex, France}

\begin{abstract}
It is well known that Raman scattering and fluorescence can be enhanced by the presence of metallic nanoparticles. Here, we derive simple equations to analyse the influence of metallic nanoparticles on upconversion processes such as non-radiative energy transfer or excited state absorption.  We compare the resulting expressions with the more familiar Raman and fluorescence cases, and find significant differences. We use numerical simulations to calculate the upconverted signal enhancement achievable by means of metallic spheres of different radii, and find particles of 100-400nm radius at infrared frequencies to be favorable. We also discuss the considerable challenges involved in using metallic particles to enhance upconversion for solar energy. 
\end{abstract}

\maketitle

\section{Introduction}

Up-conversion processes allow to obtain photons of energy significantly larger than the excitation and are present, for example, in many different rare-earths systems. Potential applications include lasers \cite{joubert99}, three dimensional imaging \cite{downing96}, quantum counters \cite{fong68} or photovoltaic cells \cite{shalav05,trupke02,conibeer07,gibart96,ivanova08}. An usual path to maximize the upconversion emission efficiency consists in trying different combinations of host matrices and rare-earth atoms. The crystal field from the host influences the transitions probabilities between the rare earth levels in a manner very dependent of the microscopic details of the quantum charge distributions. 

An alternative possibility, of interest here, relies on the possibility to locally enhance the excitation strength using metallic nanoparticles as local optical antennas. In addition, the decay rates of a given quantum emitter also depends on the environment, from purely electromagnetic, macroscopic considerations \cite{kuhn70,purcell46,drexhage70}. Metallic nanoparticles can thus be used as antennas both to enhance the incident field and to increase the emission rate. 
Surface Enhanced Raman Spectroscopy (SERS) serves as a clear demonstration of the potentiality of using such effects to obtain a strong signal. Beyond a chemical contribution, the strong local fields and radiative rates from plasmon resonances yield increases of many orders of magnitude.  The first experimental demonstration of SERS was made with an ensemble of molecules adsorbed on a rough metal surface\cite{fleischmann74}. In the 1980s, as SERS experiments mainly concerned assemblies of emitters \cite{otto92}, no quantitative agreement with the theory could be found. The main reason is that the emission depends strongly on the orientation of the molecule, its distance to the antenna, and the exact shape of the latter. Recent advances on the manipulation at nanoscale enable the study of single molecule and single nanoparticles \cite{nie97,kneipp97}.
 SERS is now routinely used for sample characterization, and it is exploited in apertureless Scanning Near Field Microscopy to obtain images with subwavelength resolution \cite{moskovits85,campion98,hartschuh03}.

A significant enhancement of the fluorescence from single molecules is also achievable\cite{thomas04,farahani05,mertens07,anger06,kuhn06}. 
A recent review on the modification of single molecule fluorescence close to a nanostructure can be found in Ref. \cite{kuhn08}. Nonetheless, the signal increase is in general much weaker than for SERS, due to a distinct light emission mechanism: Raman is a coherent process, comparable to Rayleigh scattering, but fluorescence is typically incoherent, with the population and depopulation of the different energy levels regulating photon emission.

In view of the fundamental differences between these emission mechanisms, it is not a priori obvious what will be the effect of metallic nanoparticles on upconversion processes. In 2007, Polman {\it et al.} \cite{polman07} showed that  upconversion luminescence from Er ions could be used to image surface plasmons. The influence of small metallic nanoparticles on the upconverted signal from an assembly of Er ions has recently been investigated experimentally \cite{daSilva07,rai08b}, but no theoretical model was given.

The purpose of this article is to address the issue of upconversion from a single emitter located in the near field of an individual nanoparticle. We consider two upconversion mechanisms which are often predominant in systems consisting of rare earths such as Erbium or Erbium-Ytterbium:   excited state absorption ($\excitedstateabsorption$) and non-radiative energy transfer ($\nonradiativeenergytransfer$) \cite{auzel04}. One reason for optimism in the search of large signal enhancements is the proportionality between the upconverted signal far from saturation and the square (or a higher power)\cite{pollnau00} of the light intensity. This reminds of the well-known fourth power relation between the Raman signal and the local electric field enhancement. Thus, one could expect upconversion gains from localized plasmonic resonances  as large as for SERS.  Large enhancements are, however, far from guaranteed, as both $\nonradiativeenergytransfer$ and $\excitedstateabsorption$ differ from Raman in an essential property. Like fluorescence, they are incoherent absorption-emission processes, which can be treated using rate equations, \emph{not} a coherent scattering formalism. In other words, they involve real, not virtual, transitions. 

We derive simple analytical equations to study the influence of the environment on upconversion processes. We also present similar equations for Raman scattering and fluorescence in order to compare the different processes.  Our main objective is to obtain simple equations that captures the essence of the influence of nanoantennas on these processes. Thus,  we use simplified level schemes and ignore some of the very complex physical phenomena that can enter in a real system. We illustrate the physical content of our model using numerical simulations of metallic particles of different size, at wavelengths where they exhibit localized Surface Plasmon Polariton ($\localizedplasmon$) resonances. We also briefly consider how differences on the level scheme translate into the enhancement of the upconversion signal (see Appendix B).

\section{model and notation}

\label{model}
\begin{figure}
  \includegraphics[width=1\textwidth] {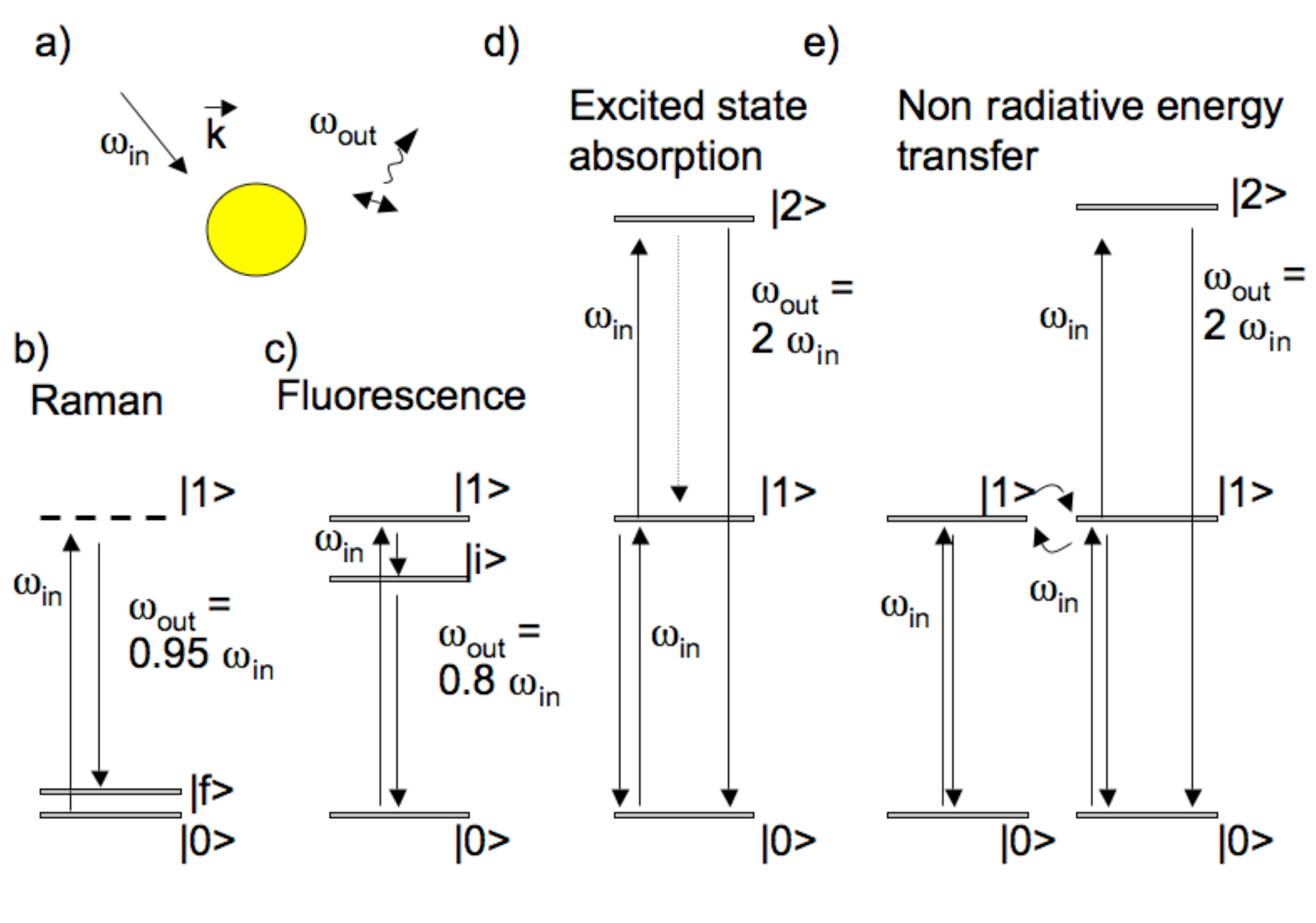}
  \caption{(a) Scheme of the general scenario considered, where the emission of a quantum system is affected by closely situated particles. The incident and emission frequencies are different. (b-e) Energy levels used in most of this paper, for (b) Raman, (c) Fluorescence and upconversion via (d) excited state absorption and (e) non-radiative energy transfer }
  \label{energylevels}
\end{figure} 

As \Fig{\ref{energylevels}}(a) illustrates, we begin by studying a very general case, how particles imbedded in an homogeneous, loss-less medium affect the light emitted by quantum emitters with discrete energy levels -- atoms, molecules, quantum dots... The energy level schemes in \Fig{\ref{energylevels}}(b-e) illustrate the different emission processes of interest, Raman, fluorescence, $\nonradiativeenergytransfer$ and $\excitedstateabsorption$. 

An experimental upconversion system can present a very complex energy level scheme, with the different transition rates --of possible electric dipolar, magnetic dipolar or even electric quadrupolar character-- very sensitive to the exact microscopic details of the sample. To avoid considering so many variables, some of them not necessarily known, we consider only dipolar electric transitions and relatively simple schemes that illustrate many of the phenomena of interest. We leave to Appendix B  a further discussion of possible changes on the level scheme. 

We study the low intensity regime so that there are no saturation effects. We focus on the influence of purely electromagnetic effects that result directly from the presence of the metallic nanoparticles. We do not consider, for example, possible changes on the crystal field due to the presence of the particle. As we consider single, fixed quantum emitters, diffusion effects \cite{auzel04} which are often important for rare earths systems do not play a role. Non-local effects on the dielectric constant can in principle increase the non-radiative decay rate and thus diminish the emitted power. They are predominant at short distances to the substrate \cite{larkin04}, but they will be significantly less important at the conditions of maximum upconversion enhancement shown in the following and we have not considered them. We also assume that the particle resonance is not modified by the presence of the rare earth emitters. 

For real systems, it will be necessary to match the particles resonances, the excitation frequency and the energy levels. To facilitate comparison over a broad range of excitation frequency $\frequencyin$, the energy difference between the ground and first excited state is always matched to the incident light, and other energy differences scale directly with this value. The excitation is assumed broad enough for population equations to be applicable and for frequency shifts \cite{morowitz69,chance75} due to the environment to be negligible, but sufficiently narrow to excite only the desired transition. The magnitude of interest is the ratio between the light power emitted in all directions at the desired frequency $\frequencyout$ with and without the particles $\intensityout/\intensityoutisolated$.

\section{Analytical expressions}

We introduce in this section the equations describing the effect of the particule on the emitted power $\intensityout/\intensityoutisolated$, for upconversion, both for $\excitedstateabsorption$ and $\nonradiativeenergytransfer$, as also for the known cases of fluorescence and Raman emission.  Appendix A presents a more complete  discussion on how to derive the equations and the corresponding assumptions.
 
\label{Analytical expressions}

\subsection{Raman scattering}

We start with a short derivation of the well-known electromagnetic contribution to SERS\cite{novotnybook}. Raman is a coherent process in which incident and scattered photons differ by a small energy amount that can be transferred, e.g. to a vibrational level of a molecule. The scattering enhancement is partly due to the  enhancement of the local intensity at the scatterer position and partly due to the enhancement of the radiation emission due to the presence of the nanoantenna (appendix A). The former is characterized by the enhancement  $ \enhancement{}{\levela\levelb}$ of the projection of the field into the dipole. The indices stands for the frequency  $\hbar\frequencyin=E_1-E_0$ of the field that matches a transition between two levels. The energy levels correspond to the excitation and emitted energy according to the simplified scheme in \Fig{\ref{energylevels}}(b). The photon emission is characterized by  the radiative rate $\Gamma^R_{1f}$ in the presence of the nanoantenna and $\Gamma^{R0}_{1f}$ without antenna, at the emission frequency $\hbar\frequencyout=E_1-E_f$.  The ratio of power emitted in all directions with and without nanoparticle is then given by 
\begin{equation}
\frac{\intensityout}{\intensityoutisolated}= \left|\enhancement{}{\levela\levelb}\right|^2\frac{\Gamma^R_{1f}}{\Gamma^{R0}_{1f}}\label{ramanexpression}.
\end{equation}

We discuss in appendix A how in occasions $\left|\enhancement{}{\levela\levelb}\right|^2\sim{\Gamma^R_{1f}}/{\Gamma^{R0}_{1f}}$. The equality holds, for example, in the particular case of a confocal set-up. If the excitation and emission frequency are sufficiently close, we finally find 

\begin{equation}
\frac{\intensityout}{\intensityoutisolated}\sim\left|\enhancement{}{\levela\levelb}\right|^4.
\end{equation}
The often discussed proportionality between the intensity of the emitted light and the fourth power of the field enhancement follows from such frequently encountered conditions.  It is a direct consequence of the coherent character of the process, \emph{not} of  the particularities of Raman, and also characterizes elastic (Rayleigh) scattering.  \\

\subsection{Fluorescence}

Fluorescence is very different from the scattering case as the former involves a real transition, and can thus be described by population equations. Introducing at this state the notation for fluorescence and up-conversion, $\enhancement{}{\levelgenerala\levelgeneralb}$ gives as before the near field enhancement along the dipole orientation between levels $\bra \levelgenerala \ket$  and $\bra \levelgeneralb \ket$. $\decay{}{\levelgenerala\levelgeneralb }{\total}$ is the total decay rate between the levels and is the sum of three contributions $\decay{}{\levelgenerala\levelgeneralb}{\total}=  \left( \decay{}{\levelgenerala\levelgeneralb}{\radiative}+\decay{}{\levelgenerala\levelgeneralb}{\nonradiative}+\decayheat{}{\levelgenerala\levelgeneralb} \right)$ that represent respectively the radiative decay resulting in photon emission, the non-radiative decay due to a transfer of energy to the particle and the non-radiative intrinsic decay due to internal transitions in the quantum emitter. Neither the process described by $\decayheat{}{\levelgenerala\levelgeneralb}$ nor  by $\decay{}{\levelgenerala\levelgeneralb}{\nonradiative}$ result in photon emission to the far field, and the energy is lost, for example as heat. The quantum yield $\quantumyield{}{\levelgenerala\levelgeneralb}{}=\decay{}{\levelgenerala\levelgeneralb}{\radiative}/(\sum_{\levelgeneralc}\decay{}{\levelgenerala\levelgeneralc}{\total})$ is the fraction of quantum emitters in level $\bra \levelgenerala \ket$ that decay directly to $\bra \levelgeneralb \ket$ by emitting a photon detectable in the far field at frequency $\frequency{}{\levelgenerala\levelgeneralb }$. The sum over levels $\levelgeneralc$ serves to include formally cases in which the quantum emitter can decay from  $\bra \levelgenerala \ket$to levels others than $\bra\levelgeneralb\ket$.  We note $\decay{}{\levelgenerala\levelgeneralb}{\totalisolated},\decay{}{\levelgenerala\levelgeneralb}{\radiativeisolated}$ and $\quantumyield{}{\levelgenerala\levelgeneralb}{\isolated}$  for the values in the homogeneous medium, in the absence of any particle ($\decay{}{\levelgenerala\levelgeneralb}{\nonradiativeisolated}$=0 by definition).

In \Fig{\ref{energylevels}}(c), a quantum emitter is excited to $\bra \levelb \ket$, from which it decays to the fluorescent level $\bra\levelintermediate\ket$ before returning to the ground state $\bra\levela\ket$. Appendix A shows that population equations allow to account for the absorption and for the radiative and non-radiative processes and to derive the population of $\bra \levelintermediate\ket$. The quantum emitter decays from this level  back to the ground state $\bra \levela \ket$ with or without photon emission, as described by $\quantumyield{}{\levelintermediate\levela}{}$. Finally, the comparison of the fluorescence signal with and without nanoparticle in the low intensity illumination regime under constant illumination yields the known equation \cite{tchenio,carminati06}

\begin{equation}
\frac{\intensityout}{\intensityoutisolated}=\left|\enhancement{}{\levela\levelb}\right|^2 \frac{\quantumyield{}{\levelintermediate\levela}{}}{\quantumyield{}{\levelintermediate\levela}{\isolated}},
\label{fluoexpression}
\end{equation}
which only depends on the near-field enhancement and the ratio between the quantum yields with and without particles. A key difference with scattering is that for each excitation of the quantum emitter to $\bra \levelb \ket$ a maximum of one photon contribute to the fluorescence. Only after the quantum emitter has decayed a new excitation and subsequent photon emission can occur, and the number of the excited molecules is always lower than the total number of quantum emitters. Such limitation do not exist in scattering processes. If $\decayheat{}{\levelintermediate\levela}=0$,  the quantum yield for homogeneous, loss-less media $\quantumyield{}{\levelintermediate\levela}{\isolated}$ is equal to one and the signal enhancement is lower than  $\left|\enhancement{}{\levela\levelb}\right|^2$, equal if the particles do not introduce non-radiative losses.

 \subsection{Up-conversion: Excited State Absorption}
 
Moving to discuss upconversion processes, we consider in this section the simple energy level schemes illustrated in \Fig{\ref{energylevels}}(d,e). In excited state absorption ($\excitedstateabsorption$), a first photon excites a quantum emitter from level $\bra \levela \ket$ to  $\bra \levelb \ket$ and a second excites this quantum emitter to $\bra \levelc \ket$. The energy of the emitted photon due to the $\bra \levelc\ket \rightarrow\bra\levela\ket$ transition is here exactly twice the corresponding value for the incident photons,  which is not necessarily the case for more complicated level schemes. 

Proceeding similarly as before, we obtain in appendix A for the upconverted signal at  $\frequencyout$ 

\begin{equation}
\frac{\intensityout}{\intensityoutisolated}=\left|\enhancement{}{\levela\levelb}\right|^4 \frac{\decay{}{\levelb\levela}{\total\isolated}}{\decay{}{\levelb\levela}{\total}}\frac{\quantumyield{}{\levelc\levela}{}}{\quantumyield{}{\levelc\levela}{\isolated}}
\label{esaexpression}
\end{equation}
If $\decay{}{\levelc\levelb }{\total}=0$, the already introduced relationship $\frac{\decay{}{\levelb\levela}{\radiative}}{\decay{}{\levelb\levela}{\radiative\isolated}} = \left|\enhancement{}{\levela\levelb}\right|^2$ is verified and there is no intrinsic losses, $\intensityout/\intensityoutisolated\leq \left|\enhancement{}{\levela\levelb}\right|^2$, equal in the absence of non-radiative decay. Under such conditions, the enhancement from $\excitedstateabsorption$ and fluorescence are similar.

 \subsection{Up-conversion: Non-Radiative Energy Transfer}

The scheme in \Fig{\ref{energylevels}}(e) represents the considered levels scheme for non-radiative energy transfer, $\nonradiativeenergytransfer$.  The  energy transfered non-radiatively from the relaxation of the two-level quantum system $\atoma$ according to the transiton $\bra \levelb\ket \rightarrow\bra\levela\ket$ serves to excite the three-level quantum system $\atomb$ from level $\bra \levelb \ket$ to $\bra \levelc \ket$. 

Solving the equations  in appendix A , we obtain for the upconverted signal from the transition $\bra \levelc\ket \rightarrow\bra\levela\ket$ of the quantum system $\atomb$:

\begin{equation}
\frac{\intensityout}{\intensityoutisolated}=\left|\enhancement{}{\levela\levelb}\right|^4 \left(\frac{\decay{\equilibrium}{\levelb\levela}{\total\isolated}}{\decay{\equilibrium}{\levelb\levela}{\total}}\right)^2\frac{\quantumyield{}{\levelc\levela}{}}{\quantumyield{}{\levelc\levela}{\isolated}}
\label{nretcomplicatedexpression}
\end{equation}
where we ignore any change in the efficiency of the non radiative energy transfer process due to the particle. We have also assumed identical field enhancement for $\atoma$ and $\atomb$ and an effective decay rate $\decay{\equilibrium}{\levelb\levela}{\total}$ for the complete system that is affected by the particle in the same way as for an individual quantum emitter.  If there is no intrinsic losses, $\atomb$ decays directly from $\bra\levelc\ket$ to $\bra\levela\ket$  and $\left|\enhancement{}{\levela\levelb}\right|^2  =\frac{ \decay{\equilibrium}{\levelb\levela}{\radiative}}{ \decay{\equilibrium}{\levelb\levela}{\radiative\isolated}}$ the presence of the particles can \emph{not} increase the emission of  upconverted photons.

We observe that the obtained expressions for the two upconverted processes are proportional to the fourth power of the near-field enhancement, similarly to the usual expression for Raman signal. However, the fourth power arises in upconversion because it is a two-photon process, while in Raman it emerges from a similar contribution from the near-field enhancement and the emission efficiency of the dipole in \Eq{\ref{ramanexpression}}. Also, the upconversion processes involve real transitions as for fluorescence. Thus, when writing population equations, non-radiative relaxation terms are introduced, and the interplay between intrinsic losses and radiative and non-radiative decay rates plays a key role. Due to these terms, we expect  weaker enhancements for upconversion than for Raman scattering. In what follows, we shall analyse this interplay by modelling rigorously the electromagnetic interaction with a metallic nanoparticle.

\section{Results}

\begin{figure}
   \includegraphics[width=0.45\textwidth] {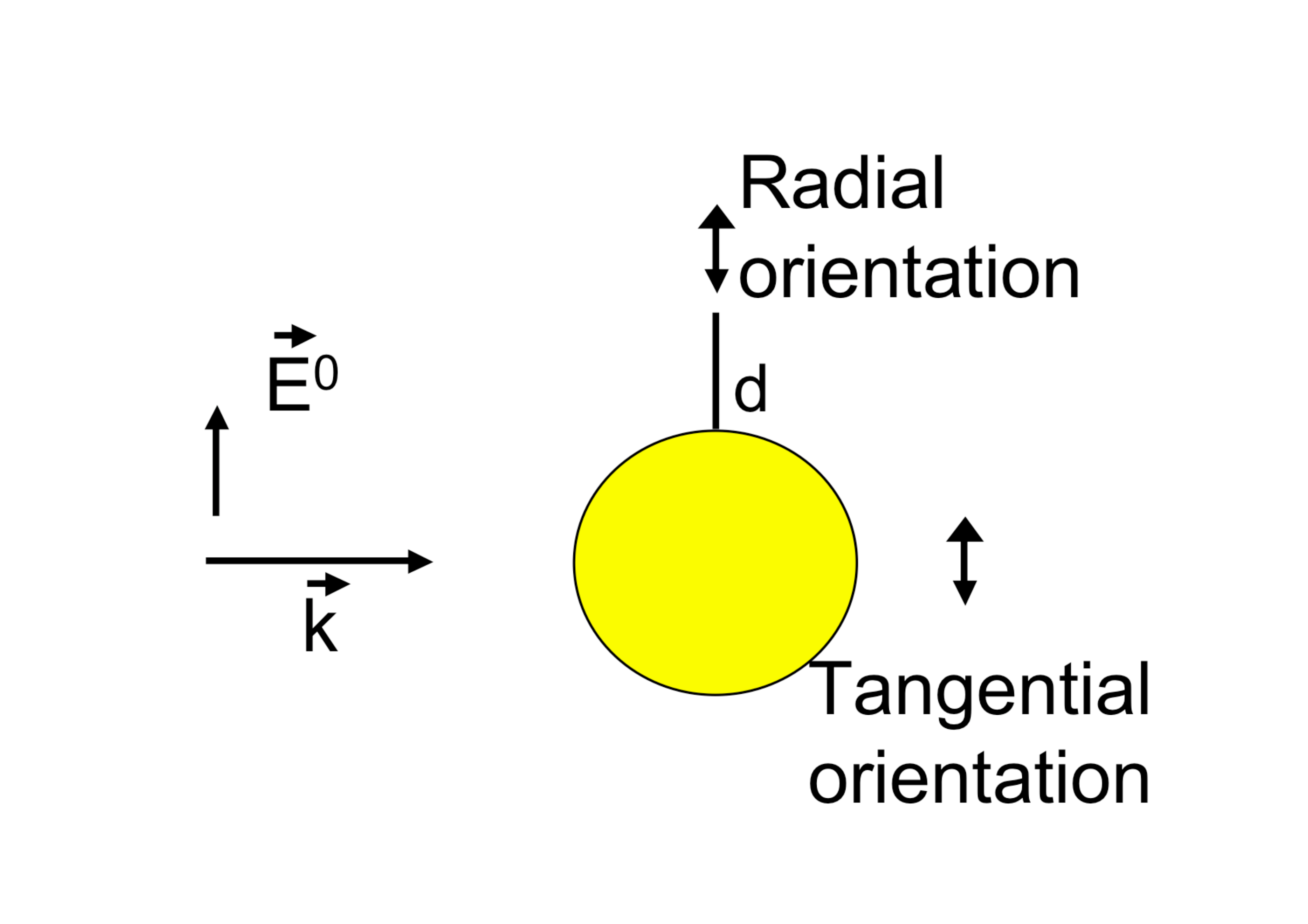}
  \caption{ A metallic nanoparticle influences the light emission from closely situated quantum systems, which can be oriented either radially or tangentially}
  \label{sphericalparticle}
\end{figure} 
We study next light emission in the proximity of a spherical metallic particle embedded in glass (with dielectric constant $\epsilonglass=2.25$). We consider (\Fig{\ref{sphericalparticle}}) dipolar excitation and emission parallel to the incident electric field vector $\evector^0$ and perpendicular to the propagation direction $\kvector$ with two orientations, one pointing toward the center of the particle (radial orientation) and the other tangential to the sphere (tangential orientation). The quantum emitter is in the plane parallel to $\evector^0$ and $\kvector$ and containing the center of the sphere, at a distance to the sphere surface $\positionmod$. For a sufficiently small particle, the total resulting electric fields are exactly  parallel to $\evector^0$ for both orientations of the dipole. For larger particles we still consider the transition to be only sensitive to the component parallel to $\evector^0$.  

We use numerical simulations  to calculate the relevant parameters entering the power emission model:  the enhancement factor and the modification of the radiative and non-radiative decay rates in presence of the particule. We then derive the modification of power radiated by the emitter in the different scenarios, Raman, Fluorescence and the two upconversion processes.

The simulations use a multiple multipole method (MAX-I)\cite{hafner99} to solve the Maxwell equations. It is a semi-analytical method where the electromagnetic field is expanded by a series of basis fields called expansions. Each of the expansions is an analytical solution of the field equations within a homogeneous domain. The decomposition of the field is numerically optimized to minimize the error at the boundary conditions. The cross-section of the quantum emitter is considered small enough not to affect the near fields of the resonant sphere.  The average error is smaller than 0.1 percent, often much smaller, and an $\sim25$ percent decrease in the number of orders of the expansions does not significantly affect the results. The dielectric constants are taken from Palik \cite{palik}.

\subsection{Decay Rates and near-field enhancement}
\label{decayenhancement}
\begin{figure}  
  \includegraphics[width=1\textwidth] {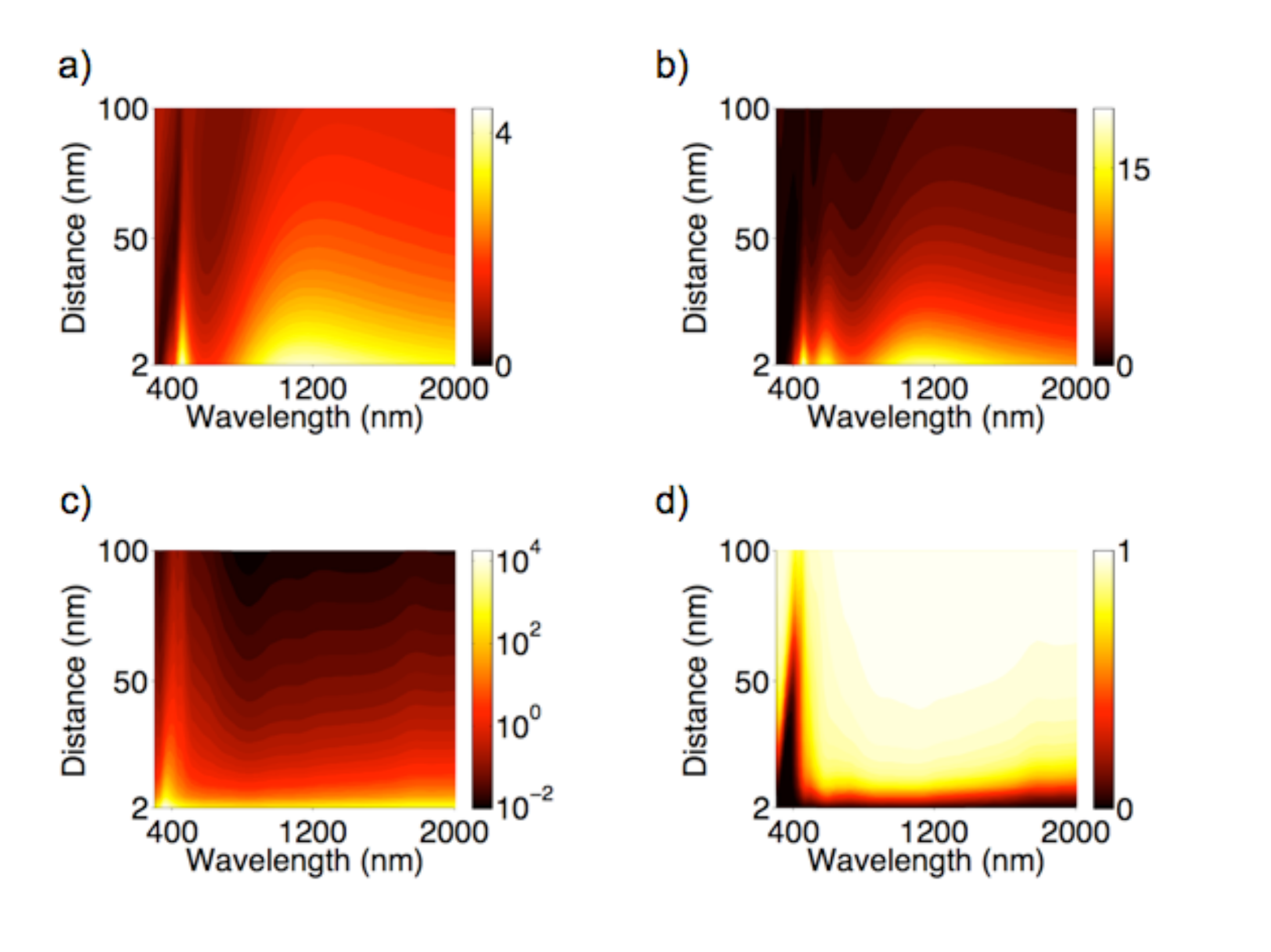}
  \caption{ Dependence as a function of the wavelength and distance to the surface for a $100\nanometer$ radius silver tip and radial orientation of (a) the field enhancement $\enhancement{}{}{}$, (b) the change on radiative decay rate $\decay{}{}{\radiative}/\decay{}{}{\radiativeisolated}$ , (c) the change on non-radiative decay rate $\decay{}{}{\nonradiative}/\decay{}{}{\radiativeisolated}$ and (d) the quantum yield for $\decayheat{}{}=0$. We note that the non-radiative decay rate (c) is plotted in logarithmic scale.}
  \label{contourelectromagnetic}
\end{figure} 
Let us first focus on the influence of both the distance to the surface $\positionmod$ and the wavelength on the parameters entering the analytical expressions for the emitted power: the field enhancement $\enhancement{}{}$ and the radiative and non-radiative rates, $\decay{}{}{\radiative}/\decay{}{}{\radiativeisolated}$ and $\decay{}{}{\nonradiative}/\decay{}{}{\radiativeisolated}$; the wavelength refers to the excitation, for the enhancement and to the emission, for the decay rates.

\Fig{\ref{contourelectromagnetic}} illustrates the distance and wavelength dependence of those parameters for the $100\nanometer$ radius silver sphere and radial orientation. It is useful for the discussion to also plot the quantum yield, with $\decayheat{}{}=0$. 
The dipolar resonance is discernible in the near-field enhancement of $\sim 4$ at around $\sim 1200 \nanometer$(\Fig{\ref{contourelectromagnetic}}(a)). This wavelength is significantly shifted to the red of the small particle value due to the non-negligible size \cite{bookkreibig}. 
We can also observe at least one higher order resonance at shorter wavelengths (around 450 nm). The change on radiative decay rate $\decay{}{}{\radiative}/\decay{}{}{\radiativeisolated}$
shown in \Fig{\ref{contourelectromagnetic}}(b) exhibits a similar behavior as the enhancement, with a broad maximum near $1200 \nanometer$; in particular, near this maximum  $\decay{}{}{\radiative}/\decay{}{}{\radiativeisolated}\sim|\enhancement{}{}|^2$ as previously discussed. At shorter wavelengths the signature of not just one, but two higher order resonances are discernible.  In contrast, the non-radiative decay rate change $\decay{}{}{\nonradiative}/\decay{}{}{\radiativeisolated}$ (\Fig{\ref{contourelectromagnetic}} (c)) behaves very differently, with a very marked maximum near $ 350 \nanometer$  up to three orders of magnitude larger than the corresponding maximum for the radiative decay rate. $\decay{}{}{\nonradiative}/\decay{}{}{\radiativeisolated}$ also increases significantly for small distances to the substrate. The conditions under which $\decay{}{}{\nonradiative}/\decay{}{}{\radiativeisolated}$ is significantly larger than $\decay{}{}{\radiative}/\decay{}{}{\radiativeisolated}$ can clearly be seen on \Fig{\ref{contourelectromagnetic}}(d) when the quantum yield tends to zero. It happens for wavelengths around $350 \nanometer$ and at small distances.

\begin{figure}  
  \includegraphics[width=1\textwidth] {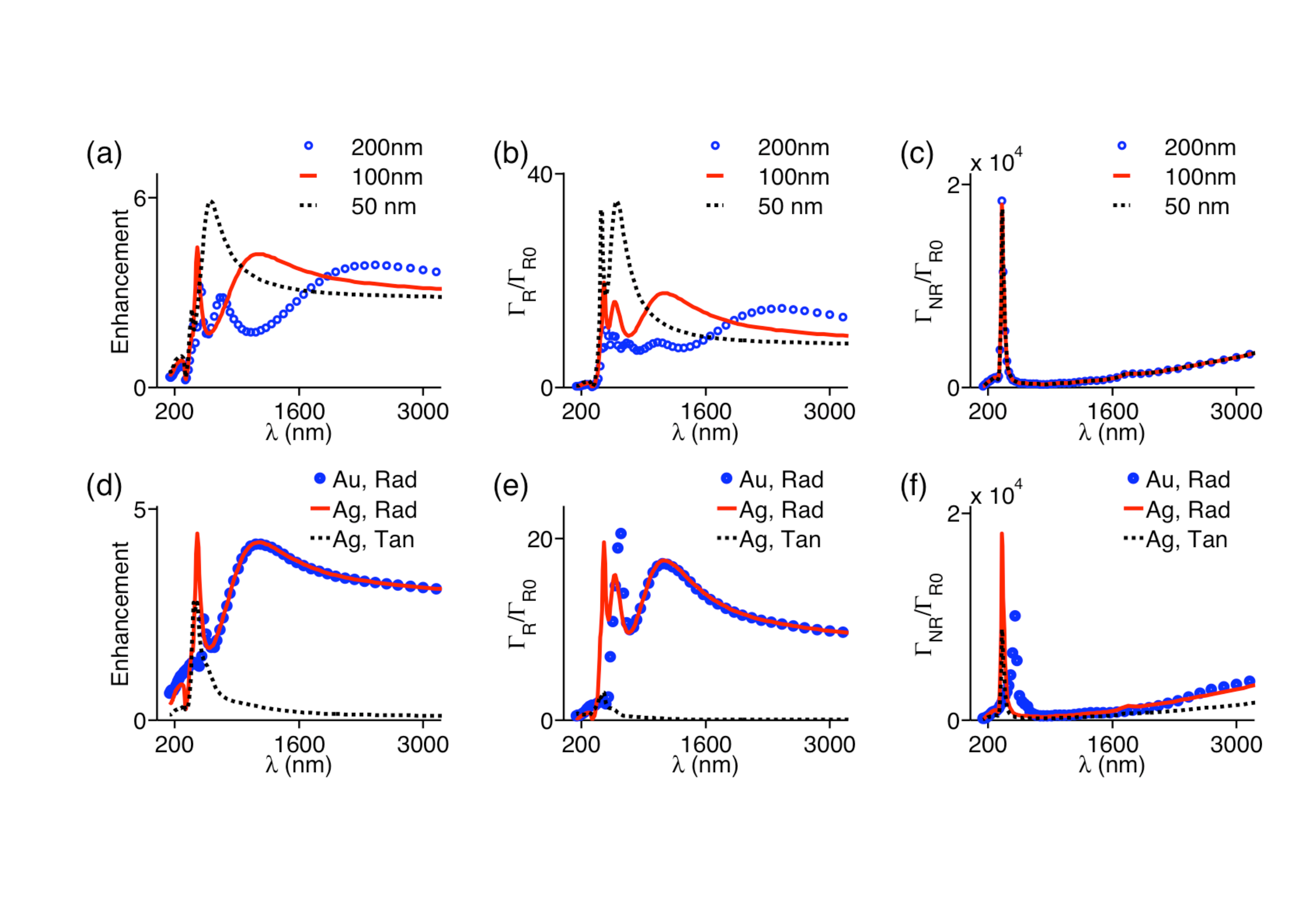}
  \caption{ Dependence with wavelength of the  (a,d) field enhancement $\enhancement{}{}{}$, (b,e) change on radiative decay rate $\decay{}{}{\radiative}/\decay{}{}{\radiativeisolated}$ and (c,f) the change on non-radiative decay rate $\decay{}{}{\nonradiative}/\decay{}{}{\radiativeisolated}$. All the plots consider a distance to the substrate of $2\nanometer$. (a-c) correspond to silver spheres of different radii and radial orientation. (d-f) correspond to 100 nm radius, for silver and gold particles and radial orientation and for tangential orientation and silver particle}
  \label{spectrumelectromagnetic}
\end{figure} 

The size of the particle can greatly influence the results. This is illustrated in \Fig{\ref{spectrumelectromagnetic}}(a-c) where the wavelength dependence for changing radius and fixed distance to the surface $\positionmod = 2\nanometer$ are studied.  If we concentrate on the dipolar resonance corresponding to the first near-field enhancement and radiative decay rate maximum starting from the longer wavelengths, we observe a shift to the red for increasing radii. Furthermore, the peak becomes weaker and broader, with the maximum near-field enhancement remaining larger than 3.5 also for $400\nanometer$ radius particles.  
The non-radiative rate change behaves again differently than the near-field enhancement as seen in  \Fig{\ref{spectrumelectromagnetic}}(c) , with the position of the maximum almost independent of size; it approximately corresponds to the real part of $\epsilonsilver$ equal to $-\epsilonglass$, the resonance for a dipole in front of a --locally-- flat surface.

The influence of the sphere material and orientation of the dipole are studied in \Fig{\ref{spectrumelectromagnetic}}(d-f) which represents the near-field enhancement and decay rates for a metallic particle of 100 nm radius of silver  with radial and tangential orientation, and for a 100 nm gold particle for radial orientation. Whereas small gold spheres are known to present a dipolar resonance redshifted with respect to silver,   \Fig{\ref{spectrumelectromagnetic}}(d-f) shows that the difference in position and strength is minimal for $100\nanometer$ radius; differences are nonetheless present for the observable high order resonances and the non-radiative decay rate peak. Finally, studying the tangential orientation for $100\nanometer$ silver particle leads to a completely different behavior of the enhancement and the radiative decay rate. Notably, the symmetry of the dipolar resonance implies strong electric fields and radiative decay rate for the radial, non tangential, orientation.

\subsection{Signal enhancement}
\label{resultsgeneral}

\begin{figure}  
  \includegraphics[width=1\textwidth] {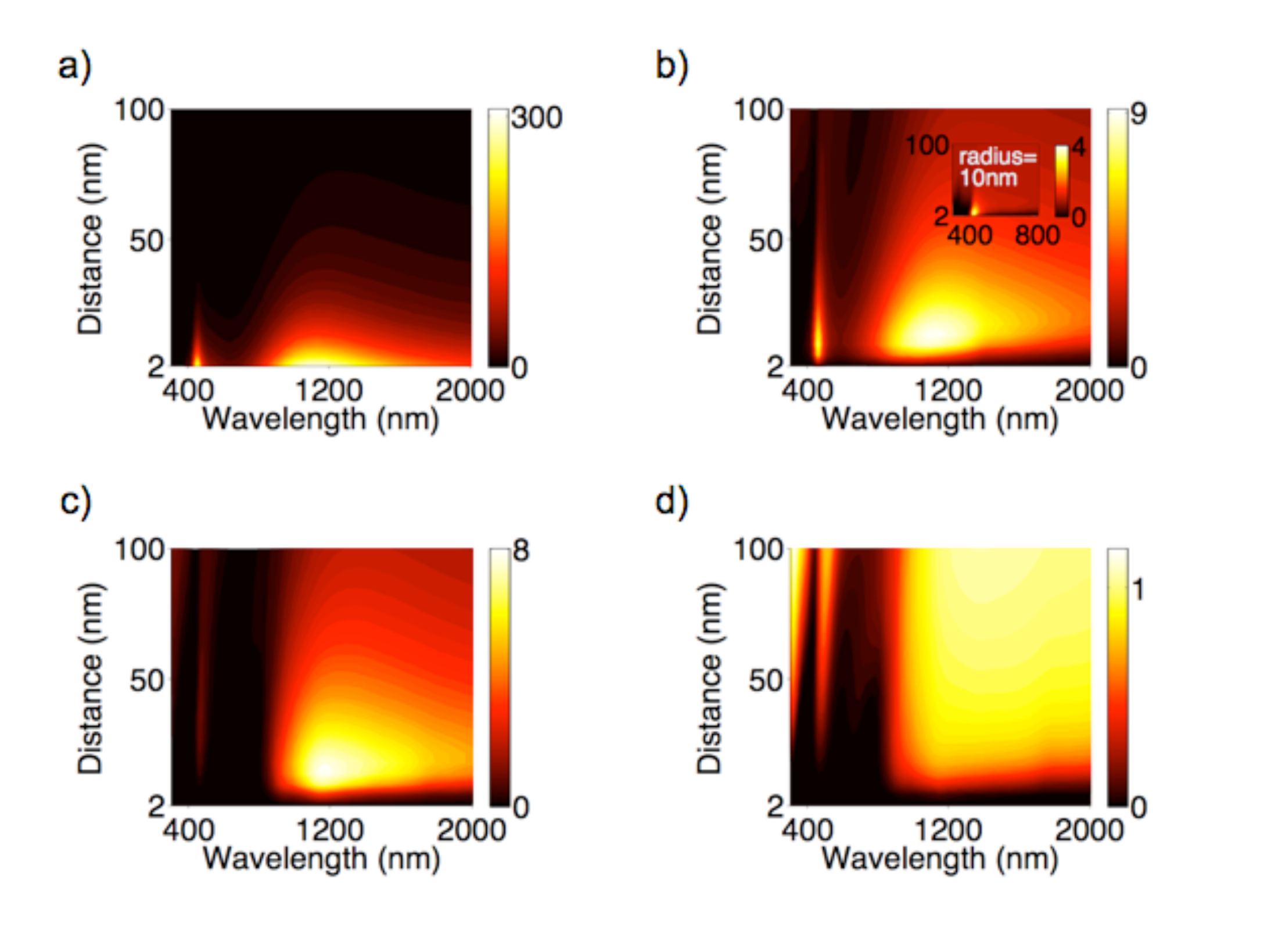}
  \caption{ Dependence as a function of the excitation wavelength and distance to the surface for a $100\nanometer$ radius silver sphere and radial orientation of the change of emitted power $\intensityout/\intensityoutisolated$ at the emission frequency of interest. (a) Raman signal, (b) fluorescence, (c) Excited state absorption and (d) Non-radiative energy transfer. The insert in (b) shows the fluorescent results for equivalent conditions but $10\nanometer$ radius.}
  \label{contourpowerout}
\end{figure}

Let us study now how the near-field enhancement and the modification of the radiative and non-radiative decay rates by the spherical particle affect the emitted power $\intensityout/\intensityoutisolated$, for the different mechanisms of Raman scattering (\Eq{\ref{ramanexpression}}), fluorescence (\Eq{\ref{fluoexpression}}), ESA upconversion process (\Eq{\ref{esaexpression}}), and NRET upconversion process (\Eq{\ref{nretcomplicatedexpression}}). To simplify, at this stage $\decayheat{}{}=0$ for all involved transitions and, for $\excitedstateabsorption$ and $\nonradiativeenergytransfer$, the decay $\bra\levelc\ket\rightarrow\bra\levelb\ket$ is negligible. To be able to study the evolution over a large wavelength range, we consider that the transition $\bra \levela \ket \rightarrow \bra \levelb\ket$ (and $\bra \levelb \ket \rightarrow \bra \levelc\ket$) coincides with the excitation frequency. The transition $\bra \levelb\ket\rightarrow\bra \levelfinal\ket$ and $\bra \levelintermediate\ket\rightarrow\bra \levela\ket$ corresponds to 0.95 and 0.8 times the excitation frequency for the Raman and the fluorescence level schemes, correspondingly. We discuss the results as a function of the excitation frequency.

We start discussing the results for a $100\nanometer$ silver sphere and radial orientation (\Fig{\ref{contourpowerout}}), if not otherwise mentioned. For Raman (\Fig{\ref{contourpowerout}} (a)), the presence of a simple metallic particle yields an enhancement of more than two orders of magnitude, with a broad maximum for the dipolar resonance and close distances to the particle. An enhancement of almost four orders of magnitude is possible at $2\nanometer$ distance for $10\nanometer$ radius. As expected, the Raman signal relates closely to the near-field enhancement (\Figs{\ref{contourelectromagnetic},\ref{spectrumelectromagnetic}}). An enhancement is also possible for fluorescence (\Fig{\ref{contourpowerout}} (b)), but significantly weaker than for Raman, because it depends on the square of the near-field enhancement, not the fourth power, and because there is a competition between this enhancement and the quenching of the emitter.

\begin{figure}  
  \includegraphics[width=1\textwidth] {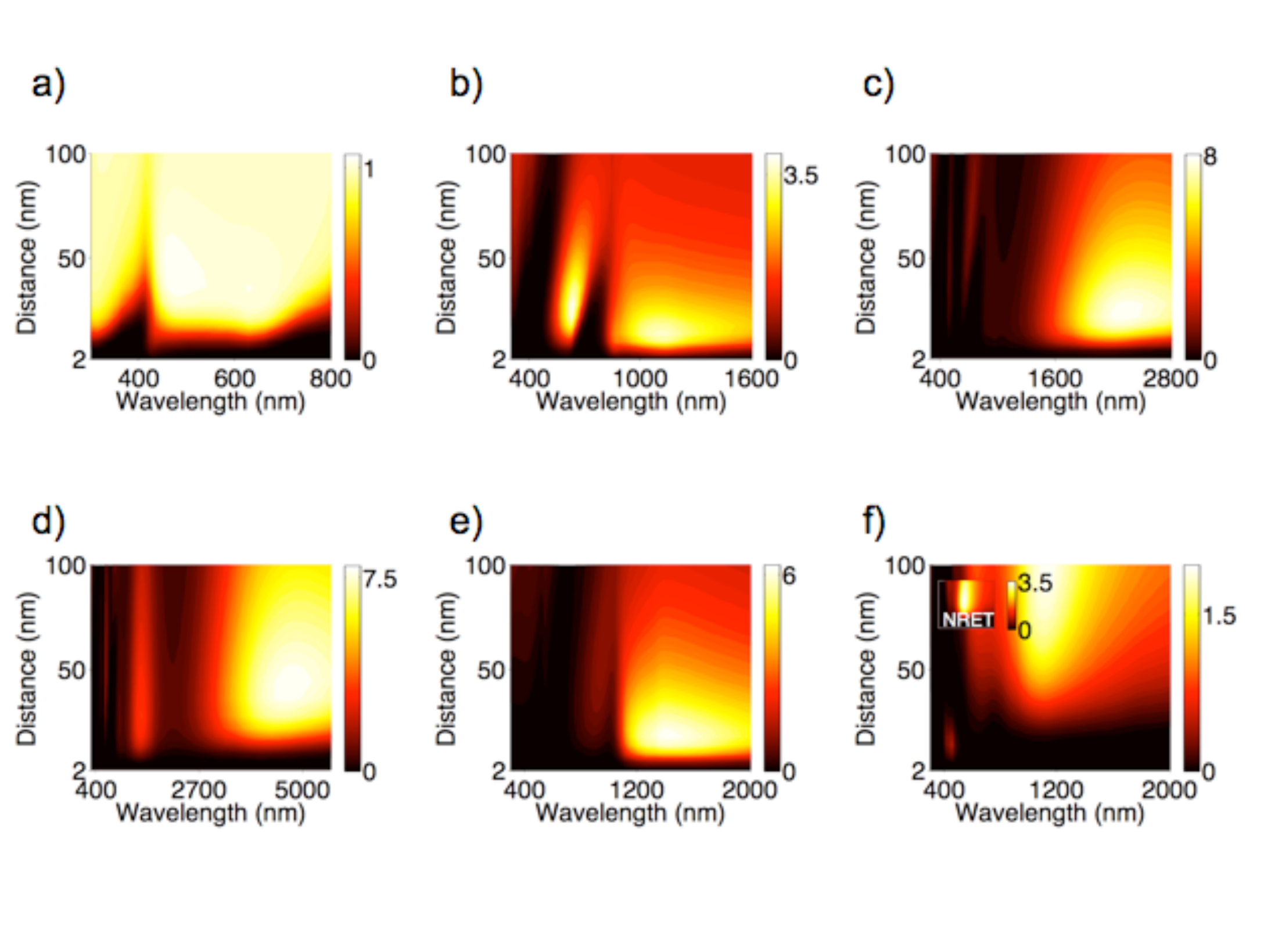}
  \caption{ Dependence as a function of the excitation wavelength and distance to the surface of the $\excitedstateabsorption$ signal $\intensityout/\intensityoutisolated$. (a) $10\nanometer$ radius, (b) $50\nanometer$ radius, (c) $200\nanometer$ and (d) $400\nanometer$ radius silver spheres and radial orientation. (e) $100\nanometer$ radius gold sphere and radial orientation.  (f) $100\nanometer$ radius silver sphere and tangential orientation. The insert in (f) represents the results for identical conditions (included the wavelength and distance range of the plot) but for $\nonradiativeenergytransfer$. Notice that the figures (a-f) use different wavelength ranges.}
  \label{contourESA}
\end{figure}

For the $\excitedstateabsorption$ case, as seen in \Fig{\ref{contourpowerout}}(c), the signal is low at small distances to the surface $\positionmod$ due to the quenching of the emitter. The signal is also specially low for excitation frequencies around $\sim 350 \nanometer$ where the non-radiative decay rate for the $\bra \levelb\ket\rightarrow\bra\levela\ket$ predominates, and at around $\sim 700 \nanometer$, which coincides with strong non-radiative decay rate and thus low quantum yield at the upconverted frequency. Related to the discussion in the previous sections, we observe that the fluorescence and  $\excitedstateabsorption$ enhancement look quite similar and both exhibit a maximum of 8-9 at the dipolar frequency. Considering a $10\nanometer$ sphere  (insert in  \Fig{\ref{contourpowerout}}(b) and \Fig{\ref{contourESA}}(a)) serves to demonstrate that the similitude is not universal.

The results are very different for NRET as seen in \Fig{\ref{contourpowerout}} (d). Indeed, the most striking feature for non-radiative energy transfer $\nonradiativeenergytransfer$ is the very small maximum signal increase (only slightly larger than 1) in the presence of the particle. The increase appears at significant distances to the surface, i.e., where the particle influence is small. The dissimilarity with ESA signal is due to $\decay{\equilibrium}{\levelb\levela}{\isolated}/\decay{\equilibrium}{\levelb\levela}{\total}$ appearing squared for $\nonradiativeenergytransfer$. Indeed, this term describes how an increase of the decay rate of $\bra\levelb\ket$ can rapidly deplete this level and diminish the probability of an excitation to $\bra\levelc\ket$, considerably affecting the upconverted signal. 

\Fig{\ref{contourESA}}(a-d) and  \Fig{\ref{contourpowerout}}(c) illustrate the influence of the particle radius, for $\excitedstateabsorption$ and radial orientation. The main effect of small silver particles, of radius below 10 nm, is to decrease the upconverted signal, which can be attributed to strong non-radiative decay rates being predominant over near-field enhancement. For the larger particles considered ($100-400\nanometer$ radius), the value of the maximum upconversion enhancement is $\sim 8$, with the position of the maximum shifting with the dipolar resonance, i.e. to the red with increasing radius. The optimal distance varies from $\sim 15\nanometer$ to $\sim 50 \nanometer$ for $100$ and $400\nanometer$ radius, respectively.  The presence of higher order resonances and the influence of the non-radiative decay rates on $\decay{}{\levelb\levela}{\isolated}/\decay{}{\levelb\levela}{\total}$ and on the $\bra\levelc\ket\rightarrow\bra\levela\ket$ quantum yield explain the presence of different secondary maxima for short wavelengths. For example, the clear minimum around $700\nanometer$ for $50\nanometer$ radius corresponds to an excitation frequency for which the non-radiative decay rate for the transition $\bra\levelc\ket\rightarrow\bra\levela\ket$ is particularly strong. 

Small spheres result in weak upconversion signal because at the position of strong enhancement, the non-radiative decay rate $\decay{}{\levelb\levela}{\nonradiative}$ can be very large. Further, the strongest near fields are obtained very close to the particle, where the quantum yield $\quantumyield{}{\levelc\levela}{}$ is also very small.  For larger spheres, the larger spacial extent of the fields and the shift of the dipolar resonance means that sufficiently small non-radiative decay rates and strong enough near fields can be simultaneously obtained. We notice also that a signal enhancement has been achieved for small metallic particles\cite{rai08,rai08b}, but for a more complex scenario that makes understanding the underlying physics challenging. We discuss in Appendix B how more complicated level schemes than here can result in upconversion enhancement also for small particles. Further, centers such as pairs of silver atoms or ions are known to influence fluorescence via non-radiative energy transfers \cite{strohhofer02,martucci05} and could also play a role in upconversion experiments.

Unsurprisingly from the results of \Section{\ref{decayenhancement}}, the $100 \nanometer$ gold spheres for radial orientation behave very similarly to their silver counterpart (\Fig{\ref{contourESA}}(e)). Last, for the $100\nanometer$ silver sphere and tangential orientation (\Fig{\ref{contourESA}}(f)), it is only possible  to obtain a weak $\excitedstateabsorption$ signal enhancement of $\sim 2$. This reminds recent studies of fluorescence which point to the convenience of radial orientation \cite{thomas04}. In contrast, for $\nonradiativeenergytransfer$, wavelengths close to $1150\nanometer$ and distances $\sim 60 \nanometer$  we obtain an enhancement of $\sim 3.5$ for tangential orientation, \emph{larger} than for radial orientation in the same conditions (\Fig{\ref{contourpowerout}}(d) and insert in \Fig{\ref{contourESA}}(f)). It is also larger than the values for $\excitedstateabsorption$ and tangential orientation (\Fig{\ref{contourESA}}(f)), which is a consequence of the term $\decay{}{\levelb\levela}{\totalisolated} / \decay{}{\levelb\levela}{\total}$ being larger than one, i.e., the total decay rate of the transition decreasing under those conditions. 

We have obtained a maximum enhancement of the upconverted signal for large particles on the $\sim 100-400 \nanometer$ radius range, for wavelengths of $\sim1$ to a few micrometers. The enhancement is about $\sim 8$ for $\excitedstateabsorption$ and radial orientation.
The maximum position is at infrared frequencies and for a distance of a few tens of nanometers to the surface. 
Significantly, for large particles the maxima for $\decay{}{}{\nonradiative}/\decay{}{}{\radiativeisolated}$ and the near-field enhancement are not at the same frequency, which allows to favor one over the other. Changing the shape of the particle, or using several particles, should allow a similar effect also for smaller dimensions\cite{bharadwaj07}. For $\nonradiativeenergytransfer$, we obtained a maximum enhancement of $\sim 3.5$ for tangential orientation.

We emphasize that this study is mostly concerned by the emission of a single quantum emitter, which could be useful, for example, for near-field imaging or to characterize individual light sources. If the objective is to use upconversion for solar applications, the question of efficiency becomes crucial. In that case, one should also consider the fact that metallic particles not only enhance the field but also absorb the incident radiation. For example, we consider an absorption cross-section between levels $\bra\levela\ket$ and $\bra\levelb\ket$ $\crosssection{}{\levela\levelb}{\absorption}=10^{-20} cm^{-2}$, and that a 2 percent of the quantum emitters in level $\bra\levelb\ket$ are excited to $\bra\levelc\ket$ and result in an upconverted photon. In the presence of the resonant spheres and ignoring saturation, an enhancement of $\sim10$ can be obtained resulting in an effective cross-section for upconversion $\sim10 \times 0.02 \times 10^{-20} =2 \,10^{-21} cm^{-2}$. This must be compared to the absorption cross-section of the particle. For a size parameter larger than one, we use the geometrical cross-section for a rough estimation of the scattering cross-section of a $100\nanometer$ radius sphere particle and obtain $\sim \pi 10^{-10} cm^{-2}$. The simulation results gives an absorption cross-section of $\sim 2.5 \,10^{-11} cm^{-2}$ at 1200nm. If we consider that all quantum emitters result in the maximum increase --no averaging over distance and orientation or considering interactions between the emitters and with the particles-- it is still necessary to have $10^{10}$ rare earths atoms per particle to reach a similar number of upconverted photons than absorbed by the particle. For an homogenous distribution between 0 and 100 $\nanometer$, an unrealistic concentration of quantum emitters of $\sim 4 \, 10^{23} cm^{-3}$ is required.

\section{Conclusion}

 To summarize, we have derived equations that give insight on how the emission from upconverted media  behaves in the presence of nanoparticles, and we have compared them with fluorescence and Raman emission. Although the emission depends on the fourth power of the local field, the detailed analysis reveals typical gains much weaker than for Raman. It is sometimes similar but can also be significantly smaller than for fluorescence. We obtain moderate enhancement of the upconverted signal when using large metallic particles illuminated in the infrared. We discuss the possibility of larger signal enhancements for more complex systems (see Appendix B). Using more complex resonant geometries can also be useful to obtain strong upconversion signal. For example, dielectric antennas can result in strong near fields without quenching \cite{laroche07}.
 
\begin{acknowledgements}

We would like to acknowledge useful discussions with Fabienne Pell{\'e}, Svetlana Ivanova and Jean-Francois Guillemoles. We also acknowledge the ANR project THRI-PV and the EU-IP ``Molecular Imaging'' (LSHG-CT-2003-503259)  for financial support. 
 
 \end{acknowledgements}
 
\section*{Appendix A : Derivation of the analytical expressions}

\subsection*{Raman scattering}

In this section, we derive the form of the enhancement of the scattered field produced by a nanoantenna. 
Let us consider a system where the excitation and the detection are produced by a source located at point $\mathbf{r}_1$. We characterize the source by an electric dipole moment $\mathbf{p}_1=p_1\mathbf{u}$ where $\mathbf{u}$ is a unit vector that describes the polarization of the source. The electric field produced at $\mathbf{r}$ reads :
\begin{equation}
\mathbf{E}(\mathbf{r})=\G(\mathbf{r},\mathbf{r}_1,\frequencyin)\mathbf{p}_1
\end{equation}
where $\G$ denotes the Green's tensor of the system at the excitation frequency $\frequencyin$. This is simply the most general linear relation between two vectors. If the emitter is in a vacuum, then the Green's tensor is the vacuum Green's tensor $\G_{0}$. If the system includes a confocal microscope or any linear optical system, the Green's tensor accounts for it. Thus, although we have introduced a dipole source, a variety of illumination conditions is included formally. In the presence of a nanoantenna, the Green's tensor can be cast in the form:
\begin{equation}
\mathbf{E}(\mathbf{r})=[\G_0(\mathbf{r},\mathbf{r}_1,\frequencyin)+\SG(\mathbf{r},\mathbf{r}_1,\frequencyin)]\mathbf{p}_1,
\end{equation}
where $\SG$ accounts for the contribution of the nanoantenna to the Green's tensor.  If we compare the field along $\mathbf{v}$ at $\mathbf{r}$ with and without the nanoantenna, we find an enhancement $K_{\frequencyin}$ given by :

\begin{equation}
K_{\frequencyin}=\frac{\mathbf{v}\cdot \G(\mathbf{r},\mathbf{r_1}, \frequencyin) \mathbf{u}}{\mathbf{v}\cdot\G_0(\mathbf{r},\mathbf{r_1}, \frequencyin) \mathbf{u}}
\end{equation}
The field induces a dipole moment along $\mathbf{v}$ given by $\mathbf{p_2}=p_2 \mathbf{v}=\mathbf{v}(\alpha\epsilon_0\mathbf{v}\cdot\mathbf{E}(\mathbf{r}))$, where $\alpha$ is the polarizability. It is either at the same frequency for Rayleigh scattering or at a slightly different frequency for Raman scattering $\frequencyout$.  The field produced by the induced dipole at a point $\mathbf{r'}$ is  given by 

\begin{equation}
\mathbf{E}(\mathbf{r'})=\G(\mathbf{r'},\mathbf{r}, \frequencyout)\mathbf{p_2}
\end{equation}
We now consider two cases. The case where we have a confocal detection system and the case where we are interested in the emission over a given solid angle. 
Let us first assume that we use a confocal detection system with both illumination and detection at the same point $\mathbf{r'}=\mathbf{r}_1$. We further assume that before detection light passes through a polarizer parallel to the unit vector $\mathbf{u}$. The signal is then proportional to 

\begin{eqnarray}
\nonumber \intensityout\propto\vert\mathbf{u}\cdot\mathbf{E}(\mathbf{r}_1)\vert^2=\vert\mathbf{u}\cdot\G(\mathbf{r}_1,\mathbf{r}, \frequencyout)\mathbf{p_2}\vert^2\\
\mathbf{p_2}=\alpha\epsilon_0\mathbf{v}(\mathbf{v}\cdot\G(\mathbf{r},\mathbf{r}_1,\frequencyin)p_1\mathbf{u})
\end{eqnarray}
We now use the reciprocity theorem that implies that $G_{mn}(\mathbf{r},\mathbf{r}_1, \frequencyout)=G_{nm}(\mathbf{r}_1,\mathbf{r}, \frequencyout)$. It follows that $\intensityout\propto\vert\mathbf{v}\cdot\G(\mathbf{r},\mathbf{r}_1,\frequencyin)\mathbf{u}\vert^2\vert\mathbf{v}\cdot\G(\mathbf{r},\mathbf{r}_1,\frequencyout)\mathbf{u}\vert^2$.  If we now compare the signal $\intensityout$ obtained in presence of the nanoantenna with the signal $\intensityoutisolated$ without nanoantenna, we get :
\begin{equation}
\frac{\intensityout}{\intensityoutisolated}=\frac{\vert\mathbf{v}\cdot\G(\mathbf{r},\mathbf{r}_1,\frequencyin)\mathbf{u}\vert^2\vert\mathbf{v}\cdot\G(\mathbf{r},\mathbf{r}_1,\frequencyout)\mathbf{u}\vert^2}
{\vert\mathbf{v}\cdot\G_0(\mathbf{r},\mathbf{r}_1,\frequencyin)\mathbf{u}\vert^2\vert\mathbf{v}\cdot\G_0(\mathbf{r},\mathbf{r}_1,\frequencyout)\mathbf{u}\vert^2} 
\end{equation}
In other words, the enhancement of the emission is equivalent to the enhancement of the illumination. When the scattering process is elastic so that $\frequencyin=\frequencyout'$, we obtain an enhancement given by 
\begin{equation}
\frac{\intensityout}{\intensityoutisolated}=\frac{\vert\mathbf{v}\cdot\G(\mathbf{r},\mathbf{r}_1,\frequencyin)\mathbf{u}\vert^4}
{\vert\mathbf{v}\cdot\G_0(\mathbf{r},\mathbf{r}_1,\frequencyin)\mathbf{u}\vert^4}=\vert K_{\frequencyin}\vert^4
\end{equation}
The equation also holds for Raman if the difference between emission and excitation frequency is sufficiently small. The exponent 4 shows that the signal varies like the fourth power of the enhancement produced by the presence of the antenna. The derivation shows that the fourth power enhancement is due to the role of the antenna that enhances both the illumination and the emission. Note that we have assumed that the detection is made at the same point and for the same polarization than the illumination. A different configuration might not yield this fourth power of the enhancement. 

In particular, we now consider that the detection system integrates over a solid angle $\Omega$. We denote $\decay{\Omega}{\omega_{out}}{\radiative}$ the emission rate of a dipole in the presence of the nanoantenna and  $\decay{\Omega}{\omega_{out}}{\radiativeisolated}$ the emission rate without the nanoantenna. The emission rate refers here to the photons emitted through a surface subtending a solid angle $\Omega$. The enhancement of the Raman process is then given by 
\begin{equation}
\frac{\intensityout}{\intensityoutisolated}=\vert K_{\frequencyin}\vert^2\frac{\decay{\Omega}{\omega_{out}}{\radiative}}{\decay{\Omega}{\omega_{out}}{\radiativeisolated}}.
\end{equation}
We have simply written here that the amplitude of the induced dipole is enhanced by a factor $K$ and that its emission in a given solid angle is enhanced by a factor $\decay{\Omega}{\omega_{out}}{\radiative}$. If we are interested in the emission over all directions, we simply use the corresponding radiative rate denoted by $\decay{}{\omega_{out}}{\radiative}$, which is the case in the text. Note that in the case that the emission pattern is not far from isotropic and is predominantly polarized as the excitation we have due to reciprocity $\vert K\vert^2\sim\frac{\decay{\Omega}{}{\radiative}}{\decay{\Omega}{}{\radiativeisolated}}$, a relationship that we have used in the text to obtain simplified equations; for $\frequencyin\sim\frequencyout$, the fourth power $\left| K_{\frequencyin} \right|^4$ is also under such conditions a good approximation for emission over all directions. 

\subsection*{Fluorescence scattering}

We present here the known derivation of the fluorescence equation\cite{tchenio,carminati06} to get familiar with the procedure followed for the upconversion expressions. As it involves a real transition, the derivation used for the Raman case is not valid. For weak illumination, the power emitted at the fluorescence frequency is due to spontaneous emission and is proportional to $N_i$, the population of the excited state. We write population equations for the different levels in order to derive the population of the excited state in stationary regime under constant illumination. The incident light excites the quantum emitter from state $\bra \levela \ket$ to state $\bra \levelb \ket$, with subsequent relaxation to $\bra \levelintermediate\ket$ before decaying back to $\bra \levela \ket$ (\Fig{\ref{energylevels}}(c)).  The different processes are absorption, radiative relaxation and non-radiative relaxation. As before, $\decay{}{\levelgenerala\levelgeneralb}{\radiative}$ describe the radiative decay rate between $\bra\levelgenerala\ket$ and $\bra\levelgeneralb\ket$,  $\decay{}{\levelgenerala\levelgeneralb}{\nonradiative}$ the non-radiative decay rate due to energy transfer to the particle and $\decayheat{}{\levelgenerala\levelgeneralc}$ the decay rate from intrinsic losses of the quantum emitter \cite{carminati06}. The total decay rate $\decay{}{\levelgenerala\levelgeneralb}{\total}$
 is the sum of the three. The absorption takes the form $\crosssection{}{\levelgenerala\levelgeneralb}{\absorption}\intensitylocal{}{\levelgenerala\levelgeneralb}={1/2\crosssection{}{\levelgenerala\levelgeneralb}{\absorption}}\sqrt{\epsilon/\mu}|\enhancement{}{\levelgenerala\levelgeneralb}|^2|\efield{}{\levelgenerala\levelgeneralb}{\isolated}\cos(\genericanglea)|^2$, with $\mu$ and $\epsilon$ the permeability and permittivity of the medium, $\crosssection{}{\levelgenerala\levelgeneralb}{\absorption}$ the absorption cross-section for the concerned transition and $\intensitylocal{}{\levelgenerala\levelgeneralb}$ the local intensity as defined by the equality. $\cos(\genericanglea)$ accounts for the projection of the field in the absence of particle $\efield{}{\levelgenerala\levelgeneralb}{\isolated}$ to the direction of the dipolar transition of interest. A possible polarization rotation due to the particle is included in $\enhancement{}{\levelgenerala\levelgeneralb}$. The relaxation is considered to be fast enough for the population of the first excited level $\bra \levelb \ket$ to be negligible. Under the weak excitation assumption so that stimulated emission can be neglected, the rate equations are:
\begin{eqnarray}
&\nonumber \population{}{\levela}+\population{}{\levelintermediate}=&\population{}{\alllevels}, \\
& \frac{\partial \population{}{\levela}}{\partial \time}=&-\frac{\population{}{\levela}\crosssection{}{\levela\levelb}{\absorption}\intensitylocal{}{\levela\levelb}}{\hbar\frequency{}{01}} +\population{}{\levelintermediate}\decay{}{\levelintermediate\levela }{\total}\label{decayfluo}.
\end{eqnarray}
  $\population{}{\levelgenerala}$ refers to the population of level $\bra \levelgenerala \ket$ and $\population{}{\alllevels}$  the population of all levels of the quantum emitter. $\population{}{\alllevels}\sim \population{}{\levela}$ for weak excitation. The first term in the right hand side of the second equation describes the excitation of a quantum emitter from the ground level and the second term the total decay rate from the intermediate to the ground level at frequency $\frequency{}{\levelintermediate\levela }$.
  
 In stationary regime,  $\partial/(\partial\time)=0$. Since,  $\intensitylocal{}{\levela\levelb} \propto |\enhancement{}{\levela\levelb}|^2$, we find from the population equations that  $\population{}{\levelintermediate} \propto |\enhancement{}{\levela\levelb}|^2/\decay{}{\levelintermediate\levela}{\total}$. The physics is clear :  in the low intensity regime, the absorption increases linearly with the enhancement of the intensity $\vert K_{01}\vert^2$ so that the population of the excited state increases also linearly. Conversely, the increased total decay rate decreases the population. When writing that the emitted power is proportional to $\Gamma_{i0}^R N_i$, we find that the power is proportional to the quantum yield $\quantumyield{}{\levelintermediate\levela}{}=\Gamma_{i0}^R/\Gamma_{i0}^T$. The comparison of fluorescence signals with and without nanoparticle finally yields the equation:

\begin{equation}
\frac{\intensityout}{\intensityoutisolated}=\left|\enhancement{}{\levela\levelb}\right|^2 \frac{\quantumyield{}{\levelintermediate\levela}{}}{\quantumyield{}{\levelintermediate\levela}{\isolated}},
\label{fluoexpression2}
\end{equation}

 \subsection*{Up-conversion: Excited State Absorption}
 
Limiting ourselves for simplicity to the simple scheme in  \Fig{\ref{energylevels}}(d), the rate equations for $\excitedstateabsorption$, under weak excitation intensity and neglecting direct excitation to level $\bra\levelc\ket$, are

\begin{eqnarray}
&\nonumber \population{}{\levela}+\population{}{\levelb}+\population{}{\levelc}=&\population{}{\alllevels} \\
& \nonumber \frac{\partial \population{}{\levela}}{\partial \time}=&-\frac{\population{}{\levela}\crosssection{}{\levela\levelb}{\absorption}\intensitylocal{}{\levela\levelb}}{\hbar\frequency{}{\levela\levelb}} +\population{}{\levelb}\decay{}{\levelb\levela }{\total}+\population{}{\levelc}\decay{}{\levelc\levela }{\total}\\
&  \frac{\partial \population{}{\levelc}}{\partial \time}=&\frac{\population{}{\levelb}\crosssection{}{\levelb\levelc}{\absorption}\intensitylocal{}{\levelb\levelc}}{{\hbar\frequency{}{\levelb\levelc}}} -\population{}{\levelc}\decay{}{\levelc\levelb }{\total}-\population{}{\levelc}\decay{}{\levelc\levela }{\total}
\label{populationesa}
\end{eqnarray}

Proceeding similarly as before, with  $\population{}{\alllevels}\sim \population{}{\levela}\gg \population{}{\levelb}\gg \population{}{\levelc}$ due to the weak excitation, and  $\intensitylocal{}{\levela\levelb} =\intensitylocal{}{\levelb\levelc} $ for our energy levels and identical polarization dependence of both transitions, we obtain for the upconverted signal at  $\frequency{}{\levelc\levela}$: 

\begin{equation}
\frac{\intensityout}{\intensityoutisolated}=\left|\enhancement{}{\levela\levelb}\right|^4 \frac{\decay{}{\levelb\levela}{\total\isolated}}{\decay{}{\levelb\levela}{\total}}\frac{\quantumyield{}{\levelc\levela}{}}{\quantumyield{}{\levelc\levela}{\isolated}}
\label{esaexpression2}
\end{equation}
In general $\quantumyield{}{\levelc\levela}{}=\decay{}{\levelc\levela}{\radiative}/(\decay{}{\levelc\levela}{\total}+\decay{}{\levelc\levelb}{\total})$, but in the results section we considered $\decay{}{\levelc\levelb}{\total}$ to be negligeable for both $\excitedstateabsorption$ and $\nonradiativeenergytransfer$. 
 
 \subsection*{Up-conversion: Non-Radiative Energy Transfer}
  Under the same assumptions than for $\excitedstateabsorption$ and the $\nonradiativeenergytransfer$ scheme in \Fig{\ref{energylevels}}(e), we can write

\begin{eqnarray}
&\nonumber \population{~\mathclap{\atomb}}{\levela}+\population{~\mathclap{\atomb}}{\levelb}+\population{~\mathclap{\atomb}}{\levelc}=\population{~\mathclap{\atomb}}{\alllevels} \\
&\nonumber \population{~\mathclap{\atoma}}{\levela}+\population{~\mathclap{\atoma}}{\levelb}=\population{~\mathclap{\atoma}}{\alllevels} \\
& \nonumber \frac{\partial\left(\population{~\mathclap{\atomb}}{\levela}+\population{~\mathclap{\atoma}}{\levela}\right)}{\partial \time}=-\frac{\population{~\mathclap{\atomb}}{\levela}\crosssection{~\mathclap{\atomb}}{\levela\levelb}{\absorption}\intensitylocal{\atomb}{\levela\levelb}}{\hbar\frequency{\,\,\,\mathclap{\atomb}}{\levela\levelb}}-\frac{\population{~\mathclap{\atoma}}{\levela}\crosssection{~\mathclap{~\atoma~}}{\levela\levelb}{\absorption}\intensitylocal{\atoma}{\levela\levelb}}{\hbar\frequency{\,\mathclap\,{\atoma}}{\levela\levelb}}+\population{~\mathclap{\atomb}}{\levelc}\decay{\atomb}{\levelc\levela}{\total}+\left(\population{~\mathclap{\atoma}}{\levelb}+\population{~\mathclap{\atomb}}{\levelb}\right)\decay{\equilibrium}{\levelb\levela}{\total}\\
&\nonumber \frac{\population{~\mathclap{\atoma}}{\levela}}{\population{~\mathclap{\atoma}}{\levelb}} \frac{\population{~\mathclap{\atomb}}{\levelb}}{\population{~\mathclap{\atomb}}{\levela}}=\frac{\partition{~\mathclap{\atoma}}{\levela}}{\partition{~\mathclap{\atoma}}{\levelb}} \frac{\partition{~\mathclap{\atomb}}{\levelb}}{\partition{~\mathclap{\atomb}}{\levela}} \exponent{\frac{-\Delta E}{kT}}=\constant \\
&  \frac{\partial \population{~\mathclap{\atomb}}{\levelc}}{\partial \time}=\upconversion{}{}{\levelb\levelc}\population{~\mathclap{\atomb}}{\levelb} \population{~\mathclap{\atoma}}{\levelb}-\population{~\mathclap{\atomb}}{\levelc}\decay{\atomb}{\levelc\levelb }{\total}-\population{~\mathclap{\atomb}}{\levelc}\decay{\atomb}{\levelc\levela }{\total}
\label{decaynret}
\end{eqnarray}
Where necessary to avoid confusion, a further index at the upper left side, as for example in $\population{~\mathclap{\atomb}}{\levelb}$, distinguish between the quantum systems $\atomb$ and $\atoma$, where $\atomb$ emits the photon at $\frequencyout$ and $\atoma$ is at the origin of the energy transfer. We have followed Page et al. \cite{page98} and assumed the two quantum systems in fast equilibrium, with an equilibrium decay rate $\decay{\equilibrium}{\levelb\levela}{\total}$ common for the the two levels $\bra\levelb\ket$ and the population relationship expressed by the fourth equality. There, $\partition{\atoma}{\levelgenerala}$ is the partition function for the level $\levelgenerala$ of $\atoma$ and $\Delta E$ (=0 in this paper) is the energy gap between the first excited level of $\atomb$ and $\atoma$. The efficiency of the non-radiative transfer to level $\bra\levelc\ket$ is proportional to the population of both $\bra\levelb\ket$ levels, with $\upconversion{}{}{\levelb\levelc}$ the proportionality constant. We assume that the particle affects the equilibrium decay rate in the same way as for a simple quantum emitter. Solving the equations and using the definition of the quantum yield as before, we obtain for the upconverted signal

\begin{equation}
\frac{\intensityout}{\intensityoutisolated}=\left|\enhancement{}{\levela\levelb}\right|^4 \left(\frac{\decay{\equilibrium}{\levelb\levela}{\total\isolated}}{\decay{\equilibrium}{\levelb\levela}{\total}}\right)^2\frac{\quantumyield{}{\levelc\levela}{}}{\quantumyield{}{\levelc\levela}{\isolated}}
\label{nretcomplicatedexpression2}
\end{equation}
where, assuming the two quantum systems sufficiently close, of same polarization dependence and with identical frequency difference between the ground and the first excited level, we have used  $\left|\enhancement{}{\levela\levelb}\right|=\left|\enhancement{\atomb}{\levela\levelb}\right|= \left|\enhancement{\atoma}{\levela\levelb}\right|$. We have also neglected any change due to the particles in the energy transfer between the two quantum systems, otherwise a term of the form $\upconversion{}{}{\levelb\levelc}/\upconversion{}{\isolated}{\levelb\levelc}$ would appear.

\section*{Appendix B : Other Scenarios}

\label{otherscenarios}

Real upconversion systems can have complicated level schemes, and their exact configuration may significantly affect the emitted signal. We illustrate now for $\excitedstateabsorption$ the consequences  of several changes on the energy level scheme of the quantum emitter.

\begin{figure}  
  \includegraphics[width=1\textwidth] {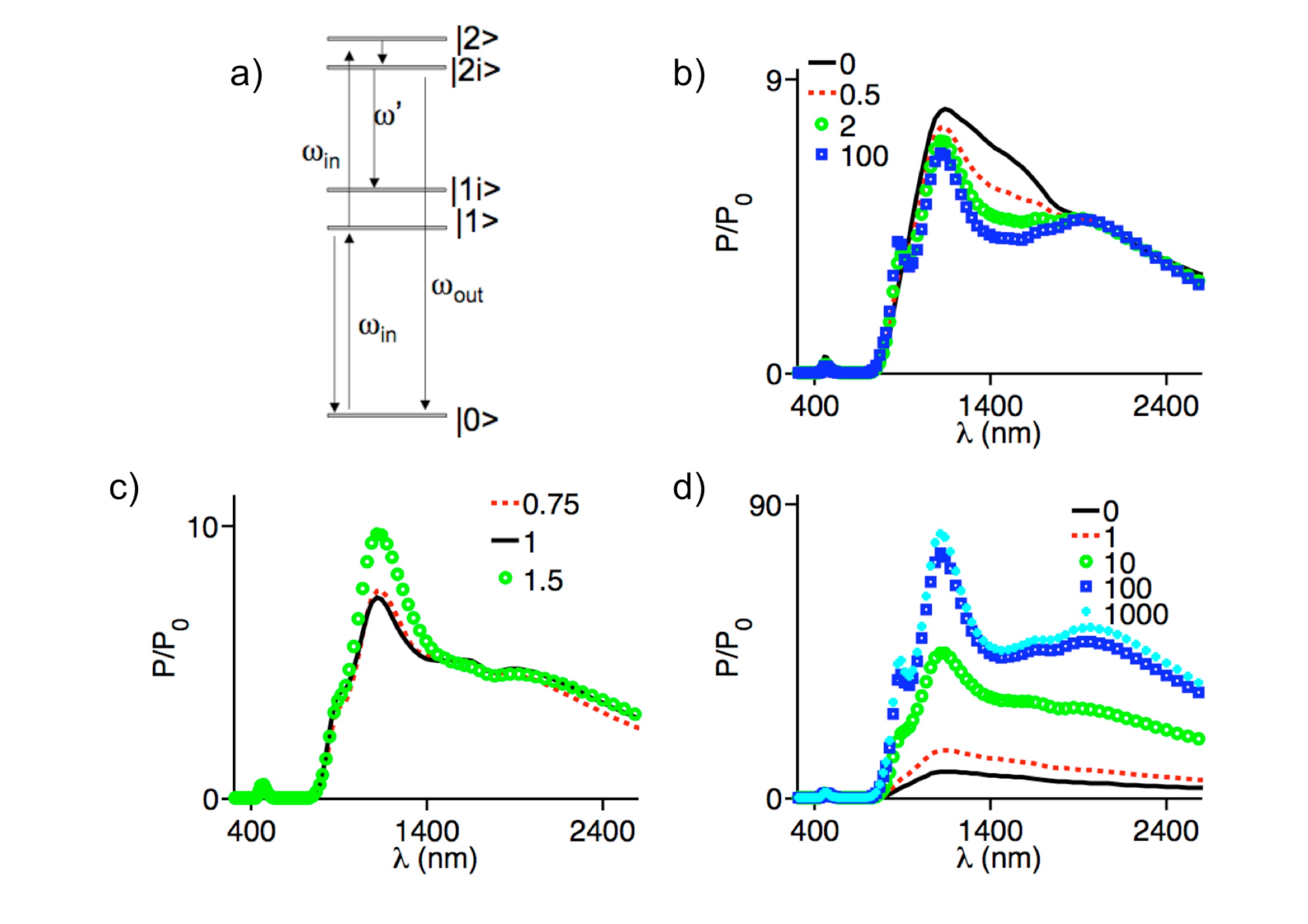}
  \caption{(a) $\excitedstateabsorption$ scheme, where the photon from $\bra\levelc\ket$ do not always decay directly to the ground level. (b-c) Excitation wavelength dependence for $100 \nanometer$ radius silver sphere, radial orientation, $15\nanometer$ distance to the surface and the level scheme in (a).  (b) Results for changing  $\decay{}{\levelintermediatec\levelintermediateb}{\radiativeisolated}/\decay{}{\levelintermediatec\levela}{\radiativeisolated}$, with $\frequencyout=1.9\frequencyin$, $\frequency{}{\levelintermediatec\levelintermediateb}=0.9\frequencyin$ and $\decayheat{}{\levelintermediatec\levelintermediateb}=\decayheat{}{\levelintermediatec\levela}=\decayheat{}{\levelintermediateb\levela}=0$. (c) Results for changing  $\frequency{}{\levelintermediatec\levelintermediateb}/\frequencyin$, with  $\frequencyout=1.9\frequencyin$, $\decay{}{\levelintermediatec\levelintermediateb}{\radiativeisolated}=\decay{}{\levelintermediatec\levela}{\radiativeisolated}$ and $\decayheat{}{\levelintermediatec\levelintermediateb}=\decayheat{}{\levelintermediatec\levela}=\decayheat{}{\levelintermediateb\levela}=0$ . (d) Results for  changing $\decayheat{}{\levelintermediatec\levela}/\decay{}{\levelintermediatec\levela}{\radiativeisolated}$, with $\frequency{}{\levelintermediatec\levelintermediateb}=0.9\frequencyin$, $\frequencyout=1.9\frequencyin$, $\decayheat{}{\levelintermediateb\levela}=\decayheat{}{\levelintermediatec\levelintermediateb}=0$ and $\decay{}{\levelintermediatec\levelintermediateb}{\radiativeisolated}=0$. }
  \label{ESAupperlevel}
\end{figure} 

\Fig{\ref{ESAupperlevel}} focuses on changes on the decay path for quantum emitters excited to level $\bra\levelc\ket$. Such quantum emitters decay fast to an intermediate level $\bra \levelintermediatec \ket$, from which it can decay to $\bra\levelintermediateb\ket$ or $\bra\levela\ket$ --emitting a photon, transferring the energy non-radiatively to the particle or losing its energy due to the intrinsic losses(\Fig{\ref{ESAupperlevel}(a)}). The particular case $\bra\levelintermediateb\ket=\bra\levelb\ket$ and $\bra\levelintermediatec\ket=\bra\levelc\ket$ reduces to the case illustrated in \Fig{\ref{energylevels}}.  For our numerical examples, if not otherwise mentioned,  $\bra\levelintermediateb\ket=\bra\levelb\ket$ and the $\bra \levelintermediatec \ket \rightarrow \bra \levelintermediateb \ket$ transition energy is 1.9 times the value for the excitation. The resulting equation looks very similar to \Eq{\ref{esaexpression}} 

\begin{equation}
\frac{\intensityout}{\intensityoutisolated}=\left|\enhancement{}{\levela\levelb}\right|^4 \frac{\decay{}{\levelb\levela}{\isolated}}{\decay{}{\levelb\levela}{\total}}\frac{\quantumyield{}{\levelintermediatec\levela}{}}{\quantumyield{}{\levelintermediatec\levela}{\isolated}}
\label{esaexpressionup}
\end{equation}
with the quantum yield of interest here taking the form

\begin{equation}
\quantumyield{}{\levelintermediatec\levela}{}=\frac{\decay{}{\levelintermediatec\levela}{\radiative}}{\decay{}{\levelintermediatec\levela}{\radiative}+\decay{}{\levelintermediatec\levela}{\nonradiative}+
\decay{}{\levelintermediatec\levelintermediateb}{\radiative}+\decay{}{\levelintermediatec\levelintermediateb}{\nonradiative}+
\decayheat{}{\levelintermediatec\levela}+\decayheat{}{\levelintermediatec\levelintermediateb}}
\label{quantumyieldup}
\end{equation}
Notice that intrinsic losses or the presence of the $\bra\levelintermediatec\ket\rightarrow\bra\levelintermediateb\ket$  always diminish the quantum yield term, but not necessarily the ratio of the quantum yield in the presence and absence of metallic particles \cite{mertens07}. Similar considerations apply to the decay $1/\decay{}{\levelb\levela}{\total}$

  \Fig{\ref{ESAupperlevel}(b)} concentrates on the influence of the strength of the decay path from level $\bra\levelintermediatec\ket$ to $\bra\levelintermediateb\ket$, with $\decayheat{}{\levelintermediatec\levelintermediateb}=\decayheat{}{\levelintermediatec\levela}=\decayheat{}{\levelintermediateb\levela}=0$ negligible for all transitions. At wavelengths around $1400\nanometer$, slightly to the red of the dipolar resonance, the decay rate from the transition $\bra\levelintermediatec\ket\rightarrow\bra\levelintermediateb\ket$ is increased more strongly than the radiative decay rate of interest $\decay{}{\levelintermediatec\levela}{\radiative}$, which negatively affects the upconverted signal. We notice that the effect is weaker for the wavelength of the maximum enhancement, because not only $\decay{}{\levelintermediatec\levelb}{\total}$ but also $\decay{}{\levelintermediatec\levela}{\radiative}$  increases significantly, the latter due to the effect of a higher order resonances. Indeed, for an adequate value of $\frequency{}{\levelintermediatec\levelintermediateb}$  the increase of $\decay{}{\levelintermediatec\levela}{\radiative}$ can boost the quantum yield with respect to the value in the absence of particle; the enhancement of the upconverted signal is then larger than if the quantum emitter always decay directly to the ground state. The achieved improvement is moderate in our example(\Fig{\ref{ESAupperlevel}(c)}).
  
To analyse the effect of the intrinsic losses from the $\bra\levelc\ket$ level, we set $\decay{}{\levelintermediatec\levelintermediateb}{\radiativeisolated}=0$ and change $\decayheat{}{\levelintermediatec\levela}/\decay{}{\levelintermediatec\levela}{\radiativeisolated}$. For particles where such intrinsic losses are considerable, an increase on the radiative decay rate  $\decay{}{\levelintermediatec\levela}{\radiative}$ due the particle can lead to a considerable larger quantum yield and subsequent signal increase(\Fig{\ref{ESAupperlevel}(d)}). 

\begin{figure}  
  \includegraphics[width=1\textwidth] {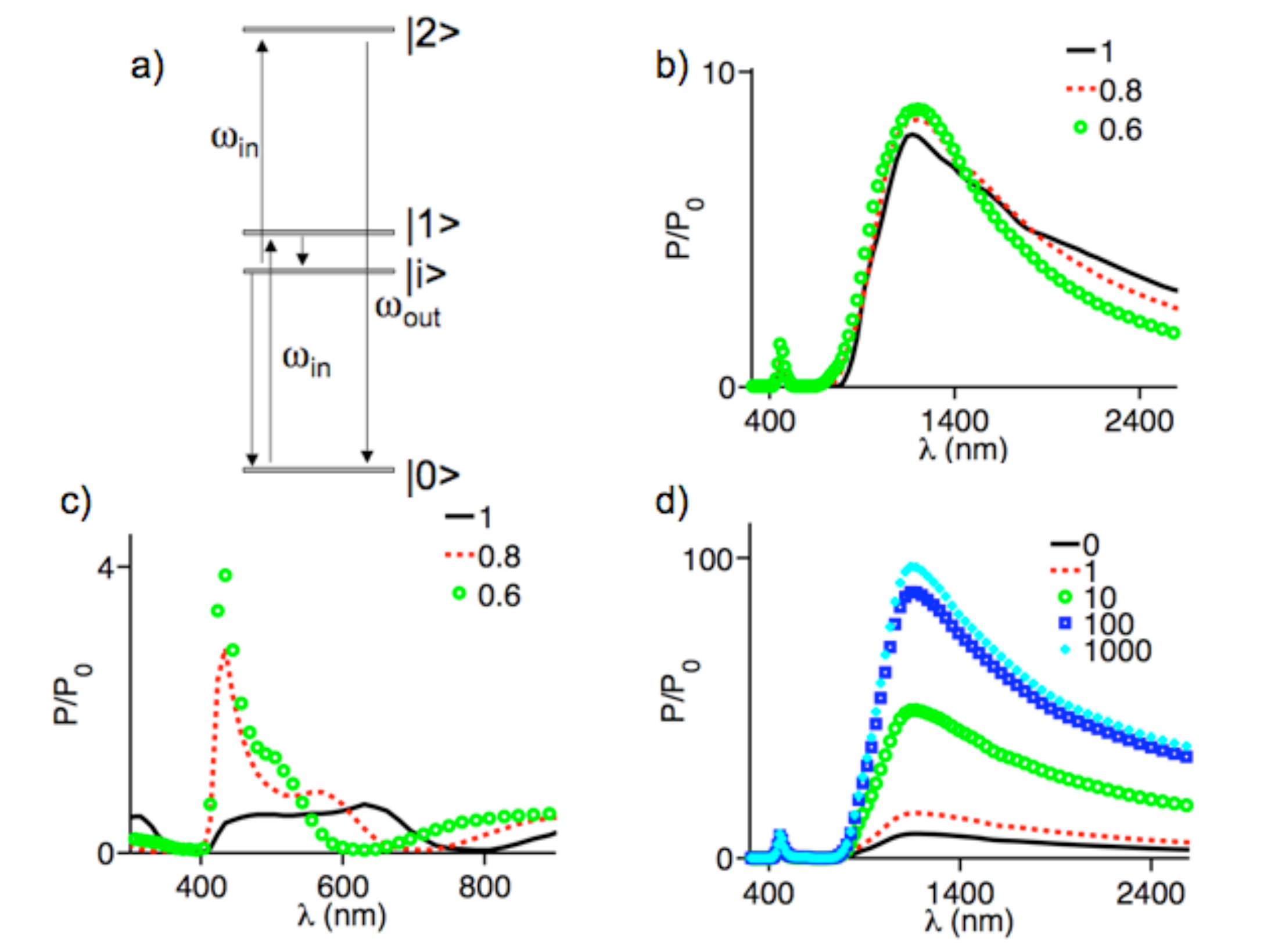}
  \caption{$\excitedstateabsorption$ scheme, where the photon from $\bra\levelb\ket$ decay non-radiatively to an intermediate state before being excited by a second photon or decaying to the ground state.(b-d) Excitation avelength dependence for silver sphere, radial orientation, $15\nanometer$ distance to the surface and the level scheme in (a).  (b) Results for changing $\frequency{}{\levelintermediate\levela}/\frequencyin$, with $\decayheat{}{\levelintermediate\levela}=\decayheat{}{\levelc\levela}=0$ and $100 \nanometer$ radius. (c) equivalent to (b) but for $10\nanometer$ radius. (d) Results for changing  $\decayheat{}{\levelintermediate\levela}/\decay{}{\levelintermediate\levela}{\radiativeisolated}$, for $\frequency{}{\levelintermediate\levela}=0.9\frequencyin$, $100 \nanometer$ radius and $\decayheat{}{\levelc\levela}=0$. }
  \label{ESAlowerlevel}
\end{figure}

\Fig{\ref{ESAlowerlevel}} centers on changes on the decay path from the first excited state. We consider that a quantum emitter excited to level $\bra\levelb\ket$ decays fast and nonradiatively to $\bra\levelintermediate\ket$ before either absorbing a second photon and being excited to $\bra\levelc\ket$ or decaying to the ground level. The resulting equation is again similar to \Eq{\ref{esaexpression}}

\begin{equation}
\frac{\intensityout}{\intensityoutisolated}=\left|\enhancement{}{\levela\levelb}\right|^4 \frac{\decay{}{\levelintermediate\levela}{\totalisolated}}{\decay{}{\levelintermediate\levela}{\total}} \frac{\quantumyield{}{\levelc\levela}{}}{\quantumyield{}{\levelc\levela}{\isolated}}
\label{esaexpressiondown}
\end{equation}
Intrinsic losses have a similar effect as before, and for large values the presence of the particle can significantly enhance the emission of upconverted signal (\Fig{\ref{ESAlowerlevel}}(d)) by increasing the term $\decay{}{\levelintermediate\levela}{\totalisolated}/\decay{}{\levelintermediate\levela}{\total}=\left(\decay{}{\levelintermediate\levela}{\radiativeisolated}+\decayheat{}{\levelintermediate\levela}\right)/\left(\decay{}{\levelintermediate\levela}{\radiative}+\decay{}{\levelintermediate\levela}{\nonradiative}+\decayheat{}{\levelintermediate\levela}\right)$. Using a $\frequency{}{\levelintermediate\levela}$ smaller than the excitation illumination allows to decouple the maximum of the total decay rate $\decay{}{\levelintermediate\levela}{\total}$ and near-field enhancement $\enhancement{}{\levela\levelb}$, and thus to boost the achievable increase on the upconverted signal. The effect for $100 \nanometer$ radius in  \Fig{\ref{ESAlowerlevel}}(b) is relatively small because for the chosen conditions and the wavelengths around the maximum the total decay rate is not a strong function of $\frequency{}{\levelintermediate\levela}$. We observe a much larger effect for $10\nanometer$ radius, \Fig{\ref{ESAlowerlevel}}(c); non-negligible enhancement of the upconverted signal becomes then possible also for small particles, in contrast to \Fig{\ref{contourESA}}(a)

\emph{Copyright 2009 American Institute of Physics. This article may be downloaded for personal use only. Any other use requires prior permission of the author and the American Institute of Physics. The following article appeared in Journal of Applied Physics and may be found at \href{http://link.aip.org/link/?jap/105/033107}{R. Esteban et al., Journal of Applied Physics 105, 033107 (2009)}.}


\end{document}